# Defect topology and annihilation by cooperative cascading movement of atoms in highly neutron irradiated graphite

Ranjan Mittal[1,2*], Mayanak K. Gupta[1], Sanjay K. Mishra[1], Sourabh Wajhal[1], Himanshu K. Poswal[3], Baltej Singh[1,2], Anil Bhimrao Shinde[1], Poluri Siva Rama Krishna[1], Peram Delli Babu[4], Ratikant Mishra[2,5], Pulya Umamaheswara Sastry[1,2], Rakesh Ranjan[6] and Samrath Lal Chaplot[1,2]

[1]*Solid State Physics Division, Bhabha Atomic Research Centre, Mumbai, 400085, India*
[2]*Homi Bhabha National Institute, Anushaktinagar, Mumbai 400094, India*
[3]*High Pressure & Synchrotron Radiation Physics Division, Bhabha Atomic Research Centre, Trombay, Mumbai 400 085, India*
[4]*UGC-DAE Consortium for Scientific Research, Mumbai Centre, R5-Shed, BARC, Trombay, Mumbai - 400 085, India*
[5]*Chemistry Division, Bhabha Atomic Research Centre, Mumbai, 400085, India*
[6]*Reactor Operations Division, Bhabha Atomic Research Centre, Mumbai, 400085, India*

**Graphite has been used as neutron moderator or reflector in many nuclear reactors. The irradiation of graphite in a nuclear reactor results in a complex population of defects. Heating of the irradiated graphite at high temperatures results in annihilation of the defects with release of an unusually large energy, called the Wigner energy. From various experiments on highly irradiated graphite samples from CIRUS reactor at Trombay and ab-initio simulations, we have for the first time identified various 2-, 3- and 4-coordinated topological structures in defected graphite, and provided microscopic mechanism of defect annihilation on heating and release of the Wigner energy. The annihilation process involves cascading cooperative movement of atoms in two steps involving an intermediate structure. Our work provides new insights in understanding of the defect topologies and annihilation in graphite which is of considerable importance to wider areas of graphitic materials including graphene and carbon nanotubes.**

Graphite has been used in high radiation environment, as neutron moderator or reflector, in many nuclear reactors, especially research and material testing reactors such as X-10 at Oak Ridge National Laboratory (USA), the Windscale Piles (UK) and G1 (Marcoule, France). About 250 kilo-tons of irradiated graphite inventory is present all over the world[1]. There is high interest in understanding of the change in behaviour of graphite, as well as in other graphitic materials such as graphene and carbon nanotubes due to irradiation[2-10]. The hexagonal structure of graphite[11] has layers of carbon atoms formed by strong covalent bonding in the a-b plane. These layers are stacked along the hexagonal axis and are held by van der Waals forces. The irradiation of graphite in a nuclear reactor results in the knocking out of carbon atoms from their equilibrium sites. Defects in graphite are unusual since they involve very large potential energy and are prevented from annealing at ambient or moderately high temperatures due to a large energy barrier. Consequently, on heating of the irradiated graphite at high temperatures of around 200 °C, the annealing of the defects is spontaneous with release of an unusually large energy, called the Wigner energy[12].

Extensive studies have been reported on unirradiated graphite, graphene and carbon nanotubes. These include studies of the highly anisotropic elastic[13, 14] and thermal-expansion behaviour[15-18] and spectroscopic studies[19-21] of the phonon spectrum using Raman, infrared and neutron scattering techniques.

The macroscopic measurements on irradiated graphite reveal change in the thermal and elastic properties due to damage in the structure[4, 12, 22-27]. Neutron irradiation damage of graphite has been studied[28] by high-resolution transmission electron microscopy and Raman spectroscopy. First-principles theoretical studies of the structure, energies and behaviour of defects in graphitic materials has also been reported[3, 29-31]. However, still there are gaps in understanding of the atomic-level structure and dynamical behavior of the knocked-out atoms, defect annealing mechanisms and the consequent Wigner energy release. Here we address these aspects.

Several graphite samples[32] irradiated with neutrons at various levels of fluence were taken out from a block of irradiated nuclear-grade graphite originally used in the reflector section of the CIRUS research reactor at Trombay, India. The highest fluence of the neutrons encountered by the samples is $2.6 \times 10^{21}$ neutrons/cm$^2$ over a period of several decades. The neutron fluence seen by various samples is depicted in **Fig. 1**. We have also used an unirradiated sample for reference. The samples have been characterized by neutron and X-ray diffraction, differential scanning calorimetry, small angle X-ray scattering, Raman scattering and specific heat measurements, and the details are given in Supplementary Information **(Supplementary Figs. S1-S6 and TABLE S1).** The results also show that the graphite samples which have been irradiated with very high neutron fluence of epithermal and fast neutrons (exceeding $10^{19}$ neutrons/cm$^2$) are damaged maximum, while the thermal neutron fluence is not so well correlated. To understand the experimental data on the structure and dynamics, we have performed ab-initio lattice dynamics and molecular dynamics simulations to model the defects, and to identify the mechanisms of annealing of defects in neutron irradiated graphite. While irradiation results in defects at various length scales, our studies have focussed on atomic level defects that are most relevant to the large Wigner energy release.

The experimental neutron diffraction data are shown in **Fig. 2a**. It can be seen that peaks in the diffraction patterns of highly irradiated graphite are broader in comparison to that in the fresh sample. These data are analyzed[33] to determine the real-space pair-distribution function, g(r), which gives the probability of finding neighbors at a distance r. **Fig. 2b** shows the pair distribution function for the unirradiated and several irradiated samples. It is evident from this figure that an additional peak in the g(r) plot of the irradiated samples appears at r=2.17 Å with a redistribution of intensity in the g(r) function. The neutron diffraction results can be understood using ab-initio simulation of the defect structure. As discussed below we find that this peak arises when an atom in the hexagonal layer is knocked-out resulting in a deformed pentagon. The peak at 2.17 Å results from one of the C-C distances in the deformed pentagon from where a vacancy has been created (**Fig. 2c**). The intensity of the peak at 2.17 Å gradually decreases with decrease of neutron fluence. This provides the experimental evidence for the defect and deformation in the hexagonal structure.

In order to study the defects in graphite we have performed simulation on a 4×4×1 supercell (comprising two graphite layers and 64 atoms) of the graphite structure. All the carbon atoms in the hexagonal a-b plane have three coordinated carbon atoms. Initially one of the carbon atoms in one of the graphite layers was moved in between the two layers, thus creating a vacancy-interstitial pair, also known as a Frenkel defect. The structural relaxation was performed for this configuration. The relaxed structure for this configuration is shown in **Fig. 2c**. It can be seen that two of the carbon atoms in the hexagon below the interstitial carbon atom form four-fold coordination due to bonding with the interstitial atom (C-C= 1.40 Å to 1.52 Å). The atoms knocked out from a hexagon results in a deformed pentagon structure, in which one of the second neighbor distances of C-C=2.45 Å (in the original hexagon) reduces to 2.06 Å (in the deformed pentagon). Three of the carbon atoms now have 2-fold coordination. The defect structure thus consists of 2-, 3- and 4-coordinated carbon atoms.

We have also simulated a 4×4×2 (128 atoms) supercell of the graphite structure with one Frenkel defect. We find the 2-, 3- and 4-coordinated carbon atoms similar to that found in the 4×4×1 supercell. We note that the small 4×4×1 supercell is able to capture the essential topological structures of the defects.

We have plotted the pair correlation function (**Fig. 2d**) in the perfect and defect structures used in our calculations. We find that the configurations with 1 Frenkel defect in 64 atoms or 128 atoms give an additional peak at about 1.5 Å, which corresponds to the four-fold coordinated carbon atoms. Further, in the g(r) plot we find additional peaks at about 2.06 Å and 2.15 Å in the defect structure with 64 and 128 atoms respectively. As discussed above, these peaks correspond (**Fig. 2c**) to one of the second neighbor C-C distances in a deformed pentagon as formed due to a vacancy of carbon atoms.

The structural studies on graphite show that defects introduced by irradiation at varying neutron fluence accompany large variation in strains. In the two simulations of one Frenkel defect in the 4×4×1 and 4×4×2 supercells, respectively, the presence of the interstitial carbon atom results in different strains along the hexagonal c-direction and only a small change along the a-direction (**Supplementary TABLE S2**). This is in agreement with our diffraction results (**Supplementary Fig. S2**).

Raman scattering is widely used to understand the nature of defects. The measured Raman spectra over 200-1800 cm$^{-1}$ from the fresh and maximum irradiated samples are shown in **Fig. 3**. Data from more samples are given in Supplementary Information (**Supplementary Figs. S3, S4**). These are in agreement with available measurements[28] above 1000 cm$^{-1}$. We further observe that in the highly irradiated sample there is a large Raman intensity over 400-1000 cm$^{-1}$ with a peak around 800 cm$^{-1}$.

It can be seen that the fresh sample shows an intense Raman mode at ~1583 cm$^{-1}$ (G mode) and a weaker peak at ~1355 cm$^{-1}$ (D mode). We note that the perfect graphite structure does not have the D mode. However, the presence of disorder in nuclear-grade graphite results in a partial breakdown of the Raman selection rules, which gives some contribution from the total phonon density of states in the Raman spectrum. Therefore, even in the fresh sample we observe small Raman intensity at around 800 cm$^{-1}$, the D mode and another small $D_o$ mode (~1620 cm$^{-1}$). For the maximum irradiated sample, the intensities of these feathers gain very significantly. For all the samples the D and G modes are found at nearly the same position at 1356 ± 4 cm$^{-1}$ and 1582 ± 4 cm$^{-1}$, respectively. The intensities of the broad low energy feather around 800 cm$^{-1}$ and the D mode, and also the peak width of both the D and G modes increase significantly for the samples irradiated with very high fluence of the epithermal and fast neutrons.

In order to understand the difference in the phonon spectrum at microscopic level we have calculated the phonon density of states as well as the in-plane and out-of-plane partial components of the phonon density of states in both the perfect and the defect structures. As discussed above, the defect structure with the 4×4×1 supercell of graphite has 2-, 3- and 4- coordinated carbon atoms within a graphite layer as well as an interstitial atom in between two layers. We have calculated the partial density of states for each of these types of carbon atoms (**Fig. 3**).

In the perfect graphite structure, the strong covalent bonding in the a-b plane results in a peak centered at 1364 cm$^{-1}$, and the out-of-plane vibrations give rise to peaks at 470 and 630 cm$^{-1}$ in the partial component along the c-axis. The Raman G-mode at ~1583 cm$^{-1}$ does not have much weight in the density of states which represents integral over the entire Brillouin zone. For the interstitial atom in the defect structure, the in-plane vibrations have a peak at about 320 cm$^{-1}$, while there is a strong peak of the out-of-plane vibrations at about 850 cm$^{-1}$ with a weaker peak at about 1300 cm$^{-1}$. The vibrational spectrum of the interstitial atom is quite different from that of the other carbon atoms within a graphite layer in both the in-plane and out-of-plane directions. This may be due to the fact that the interstitial carbon is well bonded along the c-axis with another carbon atom, while the in-layer atoms are well bonded within the a-b plane and weakly bonded along the c-axis. The out-of-plane vibrational spectra of the in-layer atoms have broad peaks at around 150-500 and 650-880 cm$^{-1}$, mainly due to disorder resulting from the removal of an atom in the a-b plane. The in-plane vibrational spectra also show broadening of peaks due to the presence of disorder. Moreover, we find remarkable changes below 500 cm$^{-1}$, where the intensity of low energy phonons is more in the spectra of the defect structure.

The changes in the observed Raman spectra (**Fig. 3**) with irradiation can be understood in terms of the calculated partial phonon density of states of the 2-, 3- and 4- coordinated carbon atoms in the defect structure. We identify (**Fig. 3**) that the most prominent increase in the intensity of the D-peak at 1360 cm$^{-1}$ is due to the increase in the 4-coordinated carbon atoms. So also, the general increase of the intensity around 800 cm$^{-1}$ may be ascribed to 2- and 3-coordinated carbon defects and the interstitial atom. The results are corroborated by simulations on a single Frankel defect in a 4×4×2 supercell (**Supplementary Fig S7**).

We have also measured the specific heat (**Fig. 4**) and observe that the specific heat, $C_P$, for the irradiated sample is more in comparison to that of the fresh graphite sample. The increase in $C_P$ is due to softening of low-frequency phonons on irradiation as discussed above.

The recombination of interstitial atoms and vacancies is the key to the release of Wigner energy in graphite. Earlier studies[29] have shown that energy of 13-15 eV is expected to be released on annihilation of a vacancy. The release of Wigner energy[12] in experiments is known to complete at about 650 K. We have verified this in our irradiated samples using Differential Scanning Calorimetry (**Supplementary Fig S5**). However, in our simulation the temperature for the defect annihilation is expected to be overestimated due to extremely small time-scale (ps) of the simulation. However, simulations are useful to understand the mechanism of annealing of defects in graphite.

The simulations are performed on the 4×4×1 supercell with one Frenkel defect at several temperatures from 300 K to 1100 K. The atomic trajectories of carbon atoms have been monitored as a function of time up to 200 ps. Simulations show that up to 800 K the interstitial carbon atom almost remains in its original position up to 200 ps. However, at 900 K within 7 ps the interstitial carbon atom at $z=0.48$ (z is the fractional coordinate along the c-axis) moved into the graphite layer at $z=0.25$. At this time, the defect structure at 900 K (**Fig. 5**) consists of five and seven-member carbon rings within the graphite layer. The structure did not relax further into hexagonal rings in the simulation up to 200 ps.

In the simulation at 1000 K, the defect energy is completely released in two steps as shown by the time dependence of the atomic coordinates and snapshot of atoms at selected times (**Fig. 5**). First at ~2.5 ps through a cooperative movement of neighboring carbon atoms we find that the interstitial carbon atom has moved closer to the vacancy. In the second step at ~5.5 ps, the hexagonal structure is restored. The simulations performed at 1100 K showed that the interstitial carbon atom at $z=0.48$ moved into the graphite layer at $z=0.25$ and within a short time of 1 ps the perfect hexagonal structure is formed. The cascading steps of cooperative movement of atoms represent the pathways of the defect annihilation process. Animations showing annihilation of defects in graphite at different temperature in various simulation cells is available in **Supplementary Information**. This provides a visualization of likely different mechanisms of annealing of defects in one or more steps as discussed above.

We found in our simulations that at a lower temperature of 900 K the defect did not fully anneal even at large time up to 200 ps. However, at 1000 K the annihilation of the defect was mediated by an intermediate step, while at 1100 K the annihilation completed in a single step.

We have calculated the energy of the defected graphite and the perfect graphite structures at various temperatures. The difference in the energy of the two structures at any temperature would give the Wigner energy released while annihilation of the defect. We find the Wigner energy at any temperature to be ~15 eV per vacancy-interstitial defect pair. However, a significant energy barrier prevents the annealing of the defect at a time-scale of 200 ps up to 800 K. However, at 900 K we found that, although the interstitial carbon atom moves in the graphite layer, the structure does not fully anneal in 200 ps time, and still retains a potential energy of about 5 eV. At 1000 K, the energy of 15 eV is released during the defect annihilation in two steps (**Fig. 6**), while at 1100 K the same energy is released in a single step.

We have also performed MD simulations with a structure consisting of two Frankel defects (**Fig. 6**) in the 4×4×1 supercell. We found that at 300 K, with two Frankel defects the stored Wigner energy is ~26.4 eV. We found that at 1000 K one Frankel defect has annealed with release of Wigner energy at ~17.5 ps, while the second Frenkel defect is not fully annealed until 100 ps as it forms a pair of five and seven-member rings. However, at 1100 K we found annealing of both the Frenkel defects in 4 ps. The annihilation of the first Frankel defect releases about 10-11 eV of energy while that of the second Frankel defect releases about 15 eV. (**See Animations in Supplementary Information**)

We have provided direct characterization of defects in neutron irradiated graphite through neutron diffraction, Raman scattering and specific heat measurements. Specifically, we have identified from the experiments and the simulations that the Frenkel defects lead to 2-, 3- and 4-coordinated carbon topologies. The microscopic understanding of annealing of defects on heating is achieved through ab-initio molecular dynamics simulations. These, in general, involve cooperative movement of atoms in several cascading steps depending on the distance between the vacancy and interstitial positions. The experimental and theoretical work has bridged the gaps in understanding of the structure and dynamical behavior of the defects in neutron irradiated graphite. Our work provides new insights in understanding of the defect annihilation in graphite and consequently release of unusually large Wigner energy.

We note from our simulations on 4×4×1 and 4×4×2 supercells that the defect structure in a graphite layer around a Frenkel defect is almost independent of the separation of such defects along the hexagonal c-axis. This may be expected due to the much weaker van der Waals interaction between the graphite layers compared to the strong covalent bonding within the layers. Other graphitic materials including graphene and carbon nanotubes have similar two-dimensional structure and bonding as in a graphite layer, and these may also be used in high radiation environment including outer space. Therefore, the present work on highly irradiated graphite is of considerable importance to wider areas of graphitic materials.

**Acknowledgements**

The use of ANUPAM super-computing facility at BARC is acknowledged. SLC thanks the Indian National Science Academy for award of an INSA Senior Scientist position.

**Author contributions**

R.M. formulated the problem and planned all the experiments and ab-initio calculations. Participated in various experiments and performed ab-initio calculations along with others. Interpretation of various experimental data and results of ab-initio calculations, and contributed in writing of manuscript.

S.L.C. formulated the problem and planned all the experiments and ab-initio calculations. Interpretation of various experimental data and results of ab-initio calculations, and contributed in writing of manuscript.

M.K.G. contributed in ab-initio calculations; interpretation of various experimental data and ab-initio calculations; and contributed in writing of manuscript. Also made animation showing annihilation of defects in graphite.

S.K.M. did X-ray diffraction, neutron diffraction and Raman scattering experiments, corresponding data analysis and contributed in writing of manuscript.

B.S. contributed in ab-initio calculations and its interpretation, and contributed in writing of manuscript.

S.W., A.B.S. and P.S.R.K. performed neutron diffraction and high-Q neutron diffraction experiments. Performed analysis of X-ray diffraction, neutron diffraction and high-Q neutron diffraction data and contributed in writing of manuscript.

H.K.P. performed Raman scattering experiments.

P.U.S. performed small angle X-ray scattering (SAXS) experiments and its analysis; and contributed in writing of manuscript.

R.K.M. performed differential scanning calorimeter (DSC) measurements and its analysis; and contributed in writing of manuscript.

P.D.B. performed specific heat measurements, analysed the data and contributed in writing of manuscript.

R.R. proposed to take up the investigation and prepared the samples used in various experiments.

**Competing interests**

The authors declare no competing interests.

**Additional information**

**Supplementary information** is available for this paper at https://doi.org/xxxxxxx

**Correspondence and requests for materials** should be addressed to R.M.

## Methods

**High-Q Neutron Powder Diffraction**

The neutron powder diffraction data on graphite samples have been recorded in the Q range of up to 15 Å$^{-1}$ on the High-Q powder diffractometer[34] at the Dhruva Reactor in Bhabha Atomic Research Centre, Trombay, India. The wavelength of the neutrons is 0.783 Å. The data has been analysed by Monte Carlo G(r) method[35] to obtain the pair distribution function.

**Raman Measurements**

The Raman spectra on graphite samples were recorded using a micro Raman spectrograph JOBINYVOUN T6400 with a 20X objective equipped with Peltier cooled CCD detector. The argon ion laser wavelength of 514.5 nm was used for measurements. The Lorentzian function was fitted to Raman profiles of various samples in the range from 1200 to 1800 cm$^{-1}$ to obtain the full-width at half-maximum (FWHM), peak intensities and peak positions of the D, G and $D_o$ modes.

**Specific Heat**

Heat Capacity was measured using heat capacity option of commercial Quantum Design make Physical Property measurement system (PPMS), which employs relaxation calorimetry technique.

**Ab-initio Calculations**

The first principles density functional theory (DFT) calculations were performed using the Vienna based ab-initio simulation package[36, 37] (VASP). The generalized gradient approximation with Perdew, Becke and Ernzerhof[38, 39] (PBE) functional implemented in VASP has been used for exchange correlation energy. The projector augmented wave method was used to incorporate the interaction between valance and core electrons. The plane wave basis set with maximum kinetic energy cutoff of 1000 eV was used. The Brillouin zone integration were performed on 20×20×2 mesh generated using the Monkhorst-Pack method[40]. The convergence criteria for total energy and forces were set $10^{-8}$ eV and $10^{-3}$ eV/Å respectively. The relaxation of the graphite structure was performed with including various vdW-DF non-local correlation functional[41-43]. We find that the relaxation performed using the optB88-vdW[43] (BO) functional scheme of the vdw-DFT method produces the best match with the experimental structure.

The structure relaxation and phonon calculations have been performed on three different configurations, one with the perfect graphite structure (space group P6$_3$/mmc) while other two configurations are with a single Frankel defect in 64 atoms supercell (4×4×1) and 128 atoms supercell (4×4×2) respectively with periodic boundary conditions. The Frenkel defect in the graphite structure was created by removing the one carbon atom from one layer and placed at a position away from the vacant site between the graphite layers. The defect structure was relaxed, which formed a metastable state with the interstitial carbon atom trapped between the layers (**Fig 2c**). The relaxed structures in all three configurations have been used to calculate the phonon density of states. The PHONON5.2[44] software has been used to generate the supercell and displaced configuration for phonon calculations. The partial density of states of individual atom has been calculated by projecting the individual atomic eigenvectors on the total eigenvector.

Ab-initio molecular dynamics (MD) simulation on the perfect and defected graphite structures has been performed on three configurations, namely, the perfect structure, the defect structures with single and two Frenkel defects in a periodic 4×4×1 supercell of 64 atoms. The total energy convergence for MD simulations was chosen to $10^{-6}$ eV; the total energy and forces were calculated using the zone-center point in the Brillouin zone. A time step of 2 fs is used. The simulations were performed for a set of temperatures from 300 K to 1100 K using NVT ensemble with Nose thermostat[45]. The simulations are performed up to 100-200 ps. For the defect structures, the starting atomic coordinates as obtained from the relaxed structure with a single or two Frankel defects is used in a 4×4×1 supercell. Here we have fixed the unit cell dimensions as obtained from the relaxation of the perfect graphite structure.

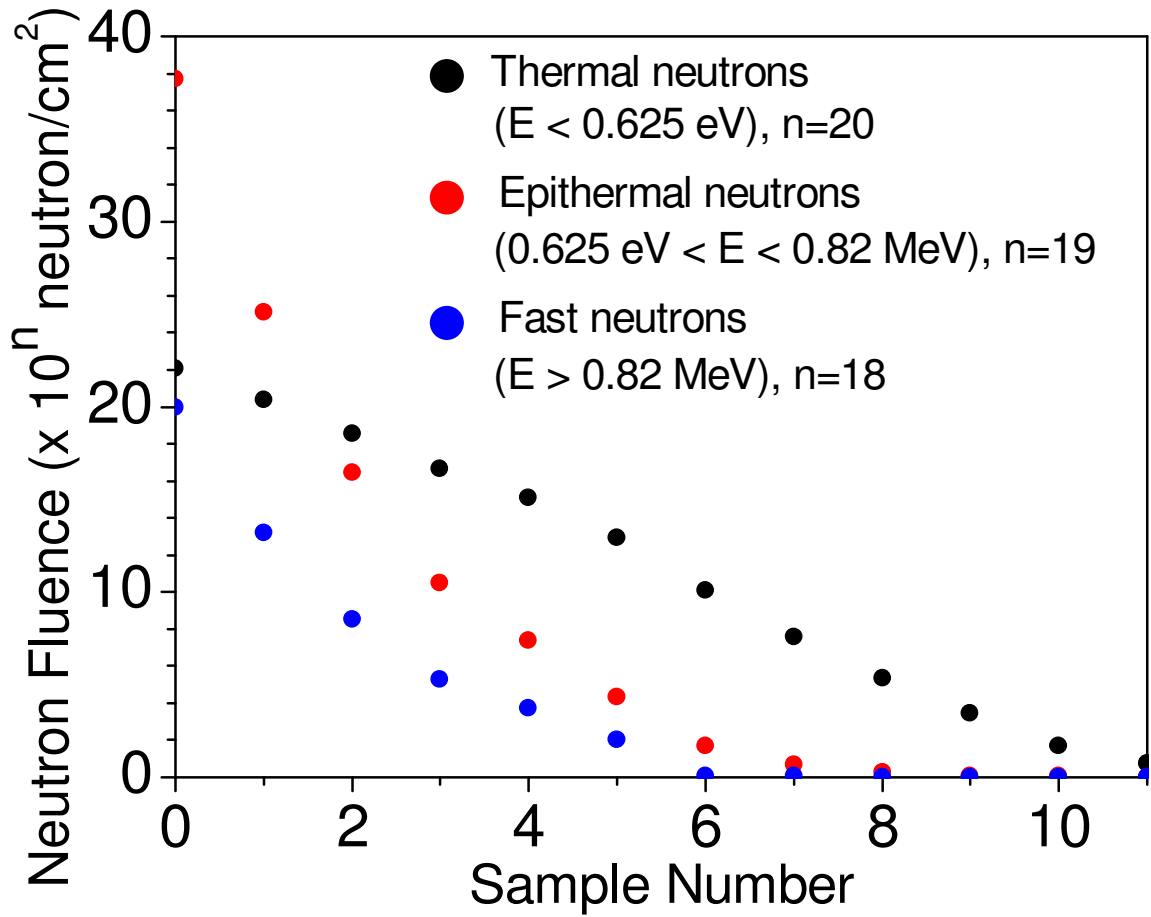

**Fig. 1 | Neutron fluence as seen by various graphite samples.** The irradiated samples are numbered as S0 to S11 in the order of decreasing neutron fluence seen by them; i.e., S0 and S11 have seen the maximum and minimum neutron fluence, respectively. Another unirradiated sample for reference is assigned as sample number S12. The unit ($10^n$ neutrons/cm$^2$) of the vertical axis is different for thermal (n=20), epithermal (n=19) and fast neutrons (n=18).

**Fig. 2 | Characterization of topological structures in defected graphite. a,** The neutron diffraction data (S(Q) vs the neutron wave-vector transfer Q) for the un-irradiated and maximum irradiated samples. **b,** The Pair distribution function of irradiated (sample S0) and un-irradiated (sample S12) graphite as obtained from powder neutron diffraction data. **c,** A graphite layer with a single Frankel defect in a 4×4×1 supercell. "*l*" and "m" correspond to the interatomic distance of 2.06 Å to 2.66 Å respectively. The interstitial atom, and the 2-, 3- and 4-coordinated carbon atoms are shown by black, blue, brown and red colors, respectively. **d,** The calculated pair correlation functions of graphite and its defect structures with a single Frankel defect in 4×4×1 and 4×4×2 supercells. In the inset, a part of the figure is zoomed.

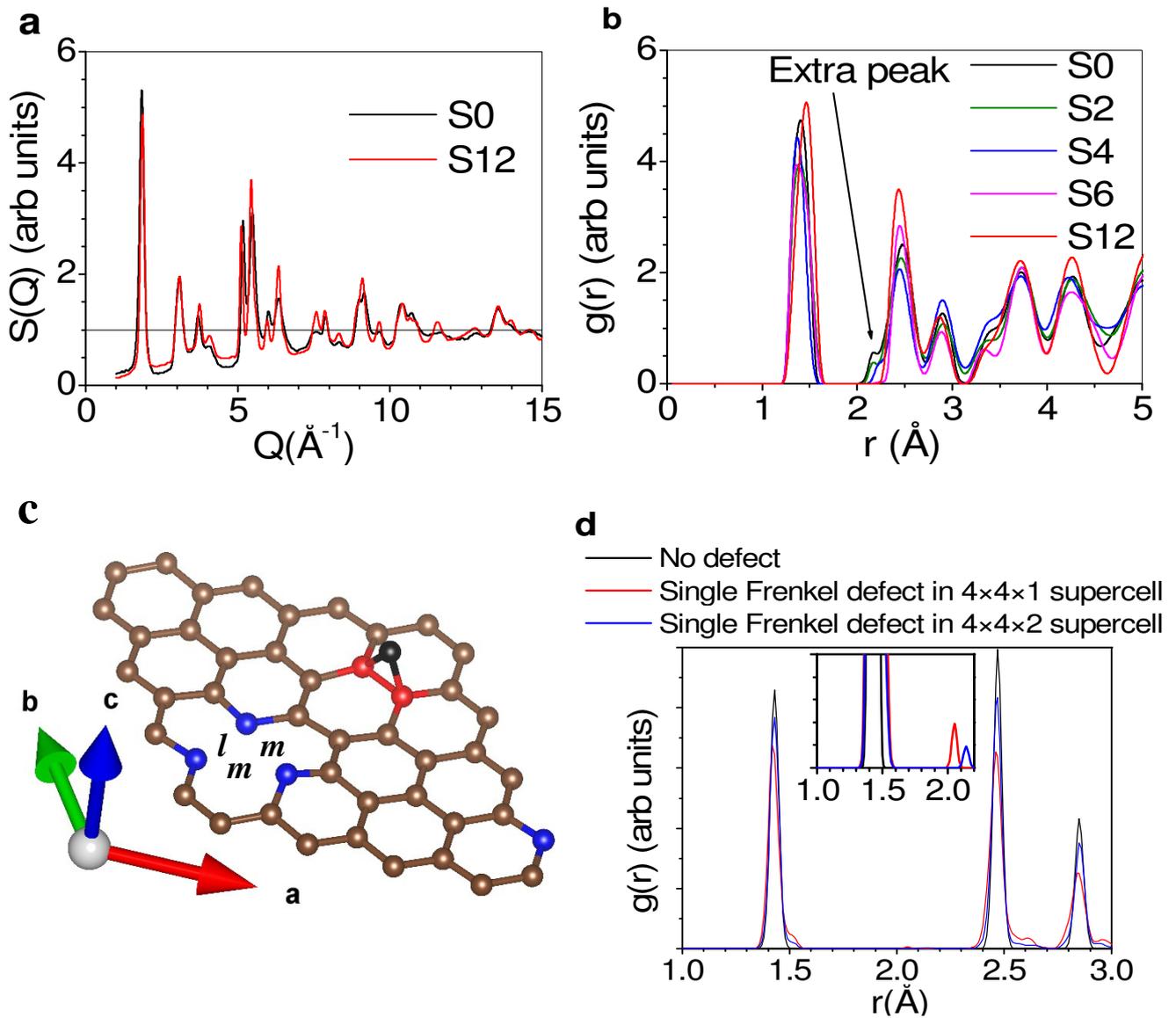

**Fig. 3 | Phonon spectra of fresh and defected graphite.** The measured Raman spectra of fresh and maximum irradiated graphite sample. The calculated partial and total phonon density of states of graphite with a single Frankel defect in a 4×4×1 supercell and with no defect. $g_x$ and $g_z$ are the x and z components of the partial phonon density of states respectively.

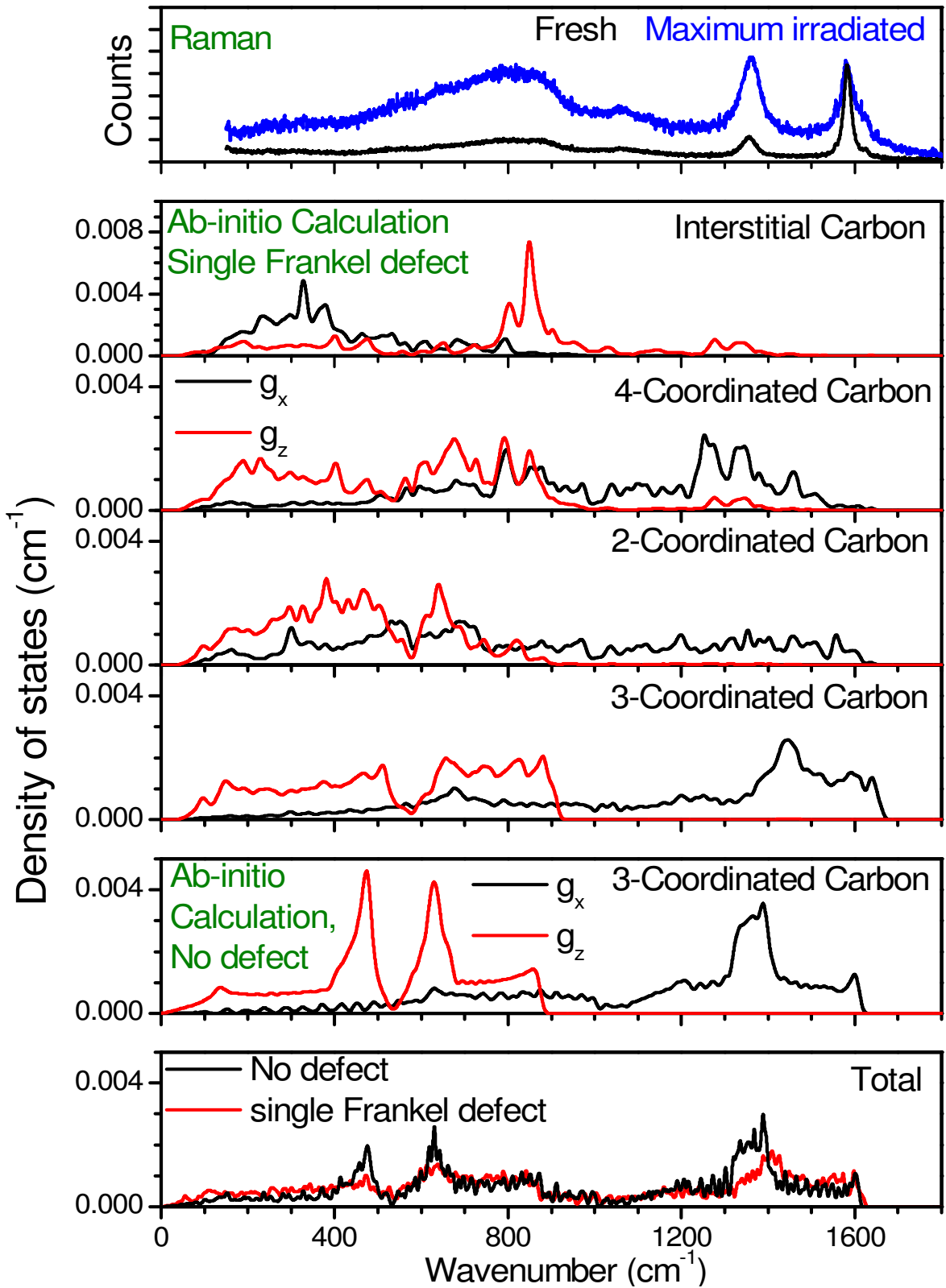

**Fig. 4 | Specific heat of fresh and irradiated Graphite.** Comparison of the measured specific heat of unirradiated (sample S12) and maximum irradiated graphite (sample S0).

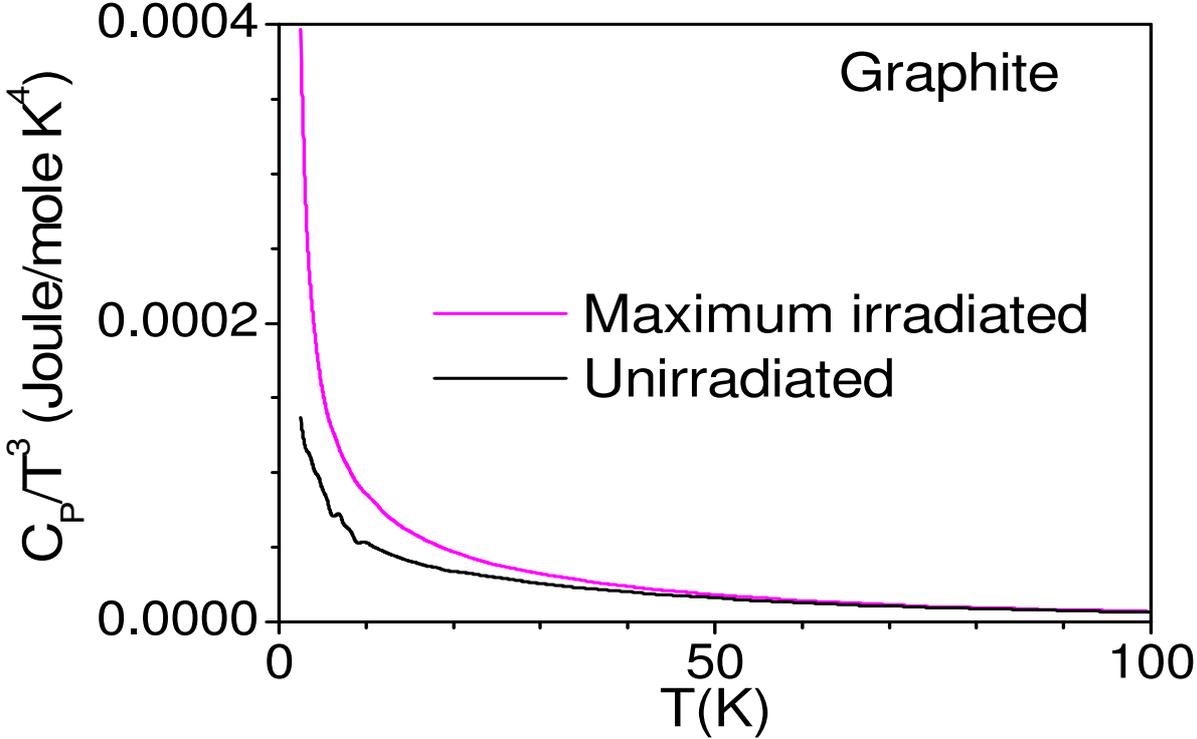

**Fig. 5 | Annihilation of defects in Graphite.** Results from ab-initio molecular dynamics simulations in a 4×4×1 supercell with a single Frankel defect at 1000 K. **a,** The time dependence of fractional coordinates of selected three carbon atoms. x, y and z are the fractional coordinate along the a-, b- and c-axis respectively. **b,** The snapshots of atoms in one layer of graphite. The defect is fully annealed in 6 ps at 1000 K. A snapshot of partially annealed defect at 900 K at 7 ps is also shown. The selected three carbon atoms identified as 1st, 6th and 11th in Figure 5a are shown in Figure 5b by black, red and green circles respectively.

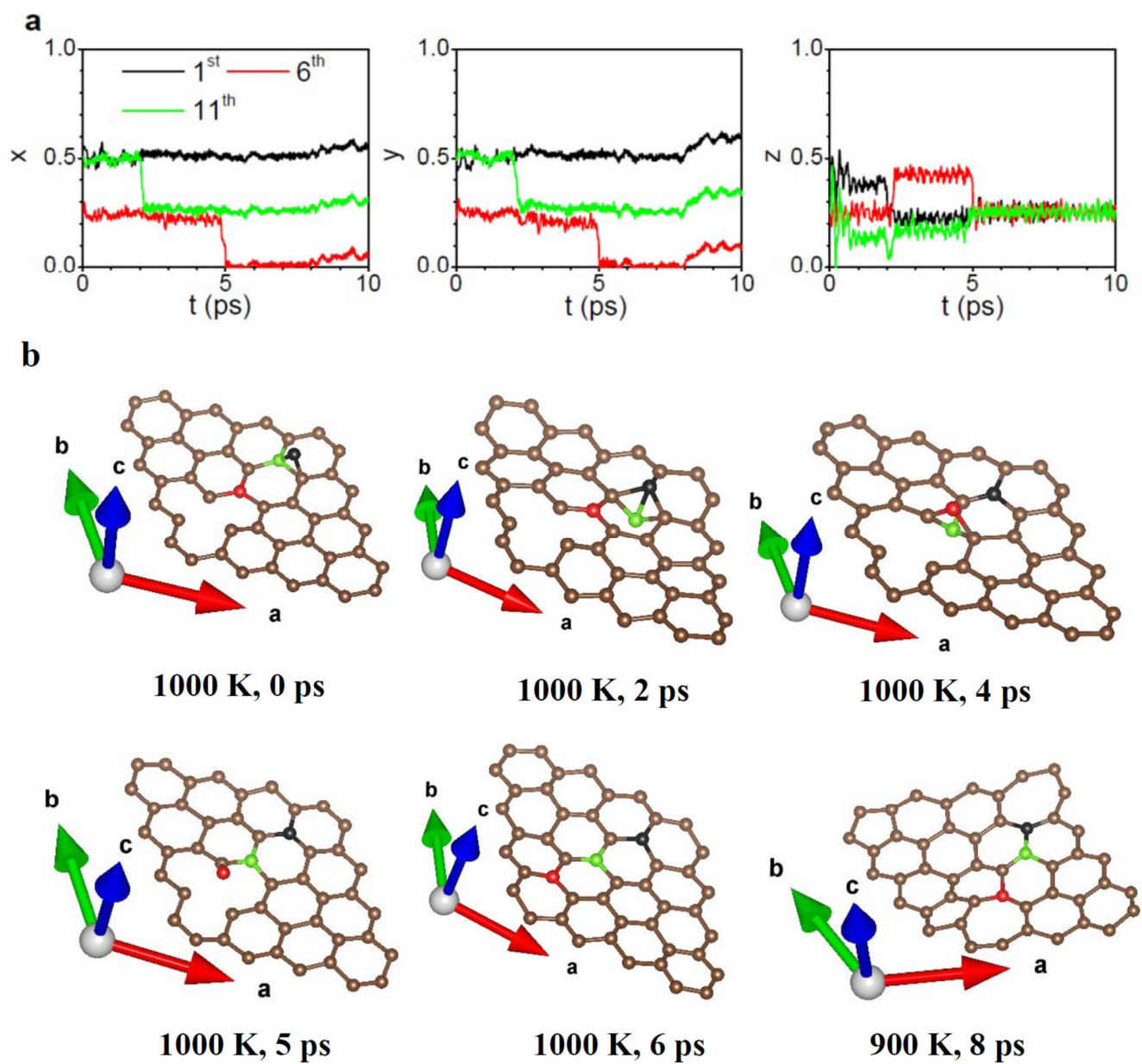

**Fig. 6 | Annihilation of defects and release of Wigner energy.** The total energy of 4×4×1 supercell (64 atoms) as a function of time with a single and two Frenkel defects from ab-initio molecular dynamics simulations. Additional results at intermediate temperatures of 700 K, 800 K and 900 K are given in Supplementary Fig S8.

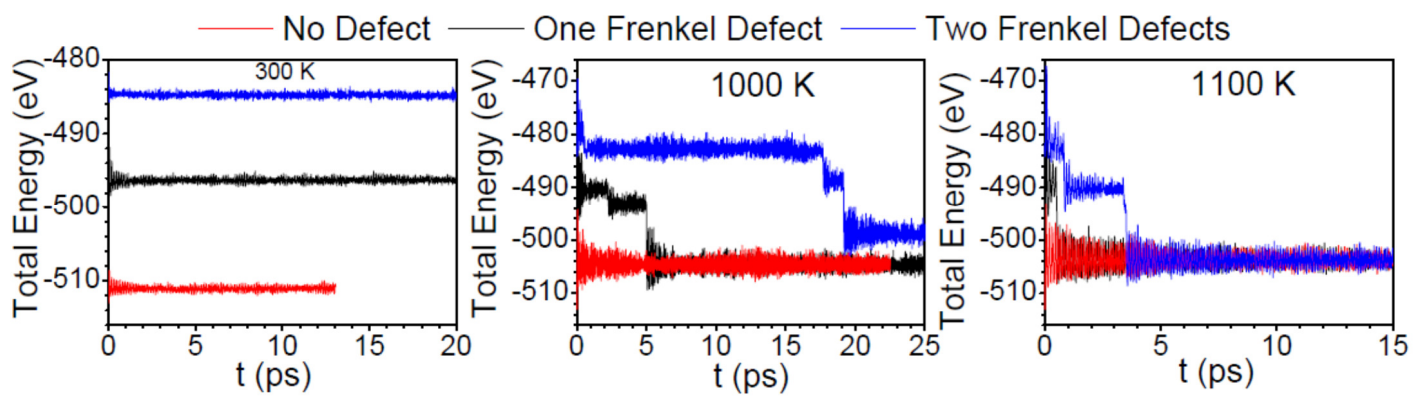

# Supplementary Information for:

# Defect topology and annihilation by cooperative cascading movement of atoms in highly neutron irradiated graphite

Ranjan Mittal[1,2], Mayanak K. Gupta[1], Sanjay K. Mishra[1], Sourabh Wajhal[1], Himanshu K. Poswal[3], Baltej Singh[1,2], Anil Bhimrao Shinde[1], Poluri Siva Rama Krishna[1], Peram Delli Babu[4], Ratikant Mishra[2,5], Pulya Umamaheswara Sastry[1,2], Rakesh Ranjan[6] and Samrath Lal Chaplot[1,2]

[1]*Solid State Physics Division, Bhabha Atomic Research Centre, Mumbai, 400085, India*
[2]*Homi Bhabha National Institute, Anushaktinagar, Mumbai 400094, India*
[3]*High Pressure & Synchrotron Radiation Physics Division, Bhabha Atomic Research Centre, Trombay, Mumbai 400 085, India*
[4]*UGC-DAE Consortium for Scientific Research, Mumbai Centre, R5-Shed, BARC, Trombay, Mumbai - 400 085, India*
[5]*Chemistry Division, Bhabha Atomic Research Centre, Mumbai, 400085, India*
[6]*Reactor Operations Division, Bhabha Atomic Research Centre, Mumbai, 400085, India*

## I. X-ray and Neutron Diffraction Measurements and Analysis of Local Structure

X-ray diffraction studies at ambient temperature are carried out using 18 KW rotating Cu-anode based powder diffractometer operating in the Bragg- Brentano focusing geometry with a curved crystal monochromator. Data were collected in the continuous scan mode in a step interval of 0.02 degree in the 2θ range of 10°–120°. The neutron powder diffraction data on graphite samples have been recorded in the 2θ range of 4°–138° in interval of 0.1° on the powder diffractometer at the Dhruva Reactor in Bhabha Atomic Research Centre, Trombay, India. The wavelength of the neutrons is 1.2443 Å. The X-ray and neutron diffraction data are analyzed using the Rietveld[1] refinement program FULLPROF[2].

The microscopic changes in the structure can be understood from diffraction studies. **Fig S1** show the evolution of the (002), (100), (101) and (004) peaks around the wave-vector transfer Q ≈ 1.84, 2.95, 3.10 and 3.70 Å$^{-1}$, respectively in the powder X-ray and neutron diffraction patterns of the irradiated graphite samples. It is evident from this figure that on increasing the neutron fluence, the peaks (002) and (004) show appreciable change in position. The peaks are shifted towards lower Q values, which suggests expansion along the *c*- direction (<001>). On the other hand, the (100) peaks show no susceptible change in the peak position and suggest insignificant change along the *a*-direction. To quantify the changes in the lattice parameters, we refined the powder neutron diffraction patterns using hexagonal structure with space group: $P6_3/mmc$. Structural parameters obtained from Rietveld refinement are shown in **Fig S2**. We find that on increasing the neutron fluence, the *c* lattice parameter shows large expansion and the *a* lattice parameter shows a small contraction. This is consistent with the nature of bonding in graphite. Graphite is known to have van der Walls bonding along the c-axis, which is much weaker than the covalent bonding in the a-b plane.

**Figure S1:** Evolution of the powder X-ray and neutron diffraction pattern of un-irradiated and irradiated graphite. The irradiated samples are numbered as S0 to S11 in the order of decreasing neutron fluence seen by them; i.e., S0 and S11 have seen the maximum and minimum neutron fluence, respectively. Another un-irradiated sample for reference is assigned as sample number S12. For clarity, refer Fig. 1. The intensity of the left panels is zoomed for clarity.

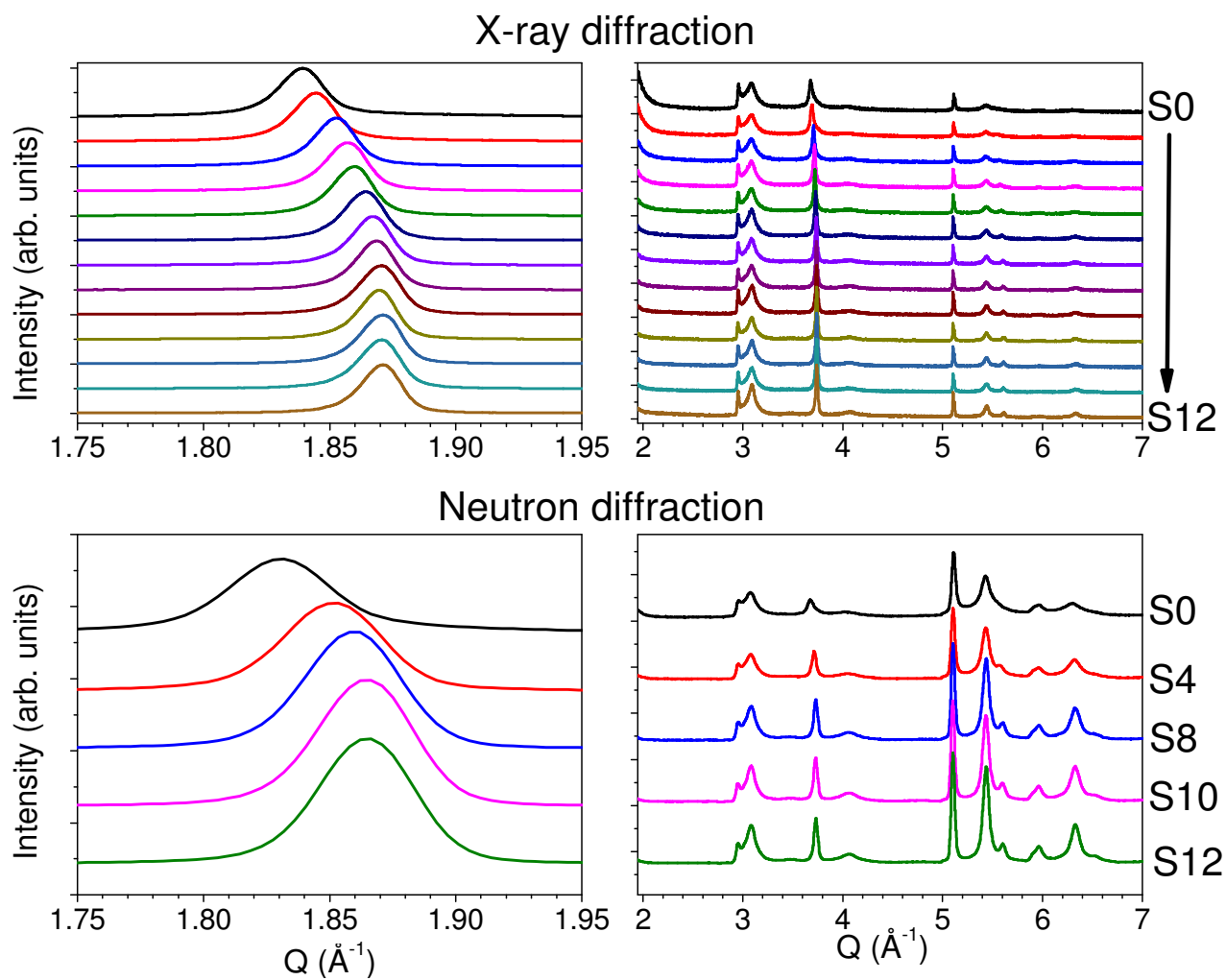

**Figure S2:** Variation of the structural parameters of graphite as obtained after Rietveld analyses of X-ray (Red coloured closed circles) and combined X-ray and neutron diffraction (Blue coloured closed circles) data with increasing fluence. Errors are within symbol size. The irradiated samples are numbered as S0 to S11 in the order of decreasing neutron fluence seen by them; i.e., S0 and S11 have seen the maximum and minimum neutron fluence, respectively. Another un-irradiated sample for reference is assigned as sample number S12. For clarity, refer Fig. 1.

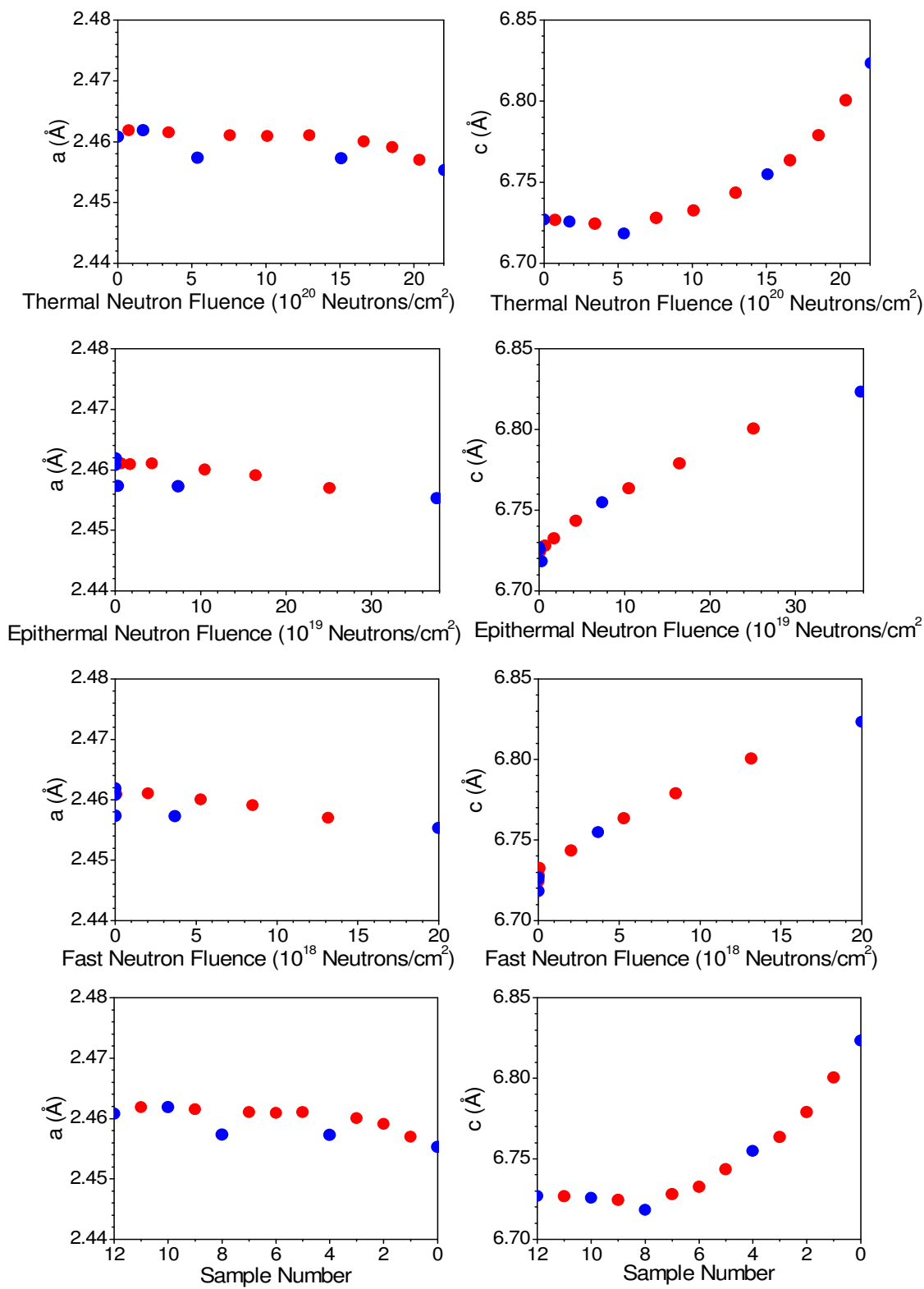

## III. Raman Scattering Measurements

**Figure S3:** Measured Raman spectra from graphite samples irradiated at different levels of neutron fluences. The irradiated samples are numbered as S0 to S11 in the order of decreasing neutron fluence seen by them; i.e., S0 and S11 have seen the maximum and minimum neutron fluence, respectively. Another un-irradiated sample for reference is assigned as sample number S12. For clarity, refer Fig. 1.

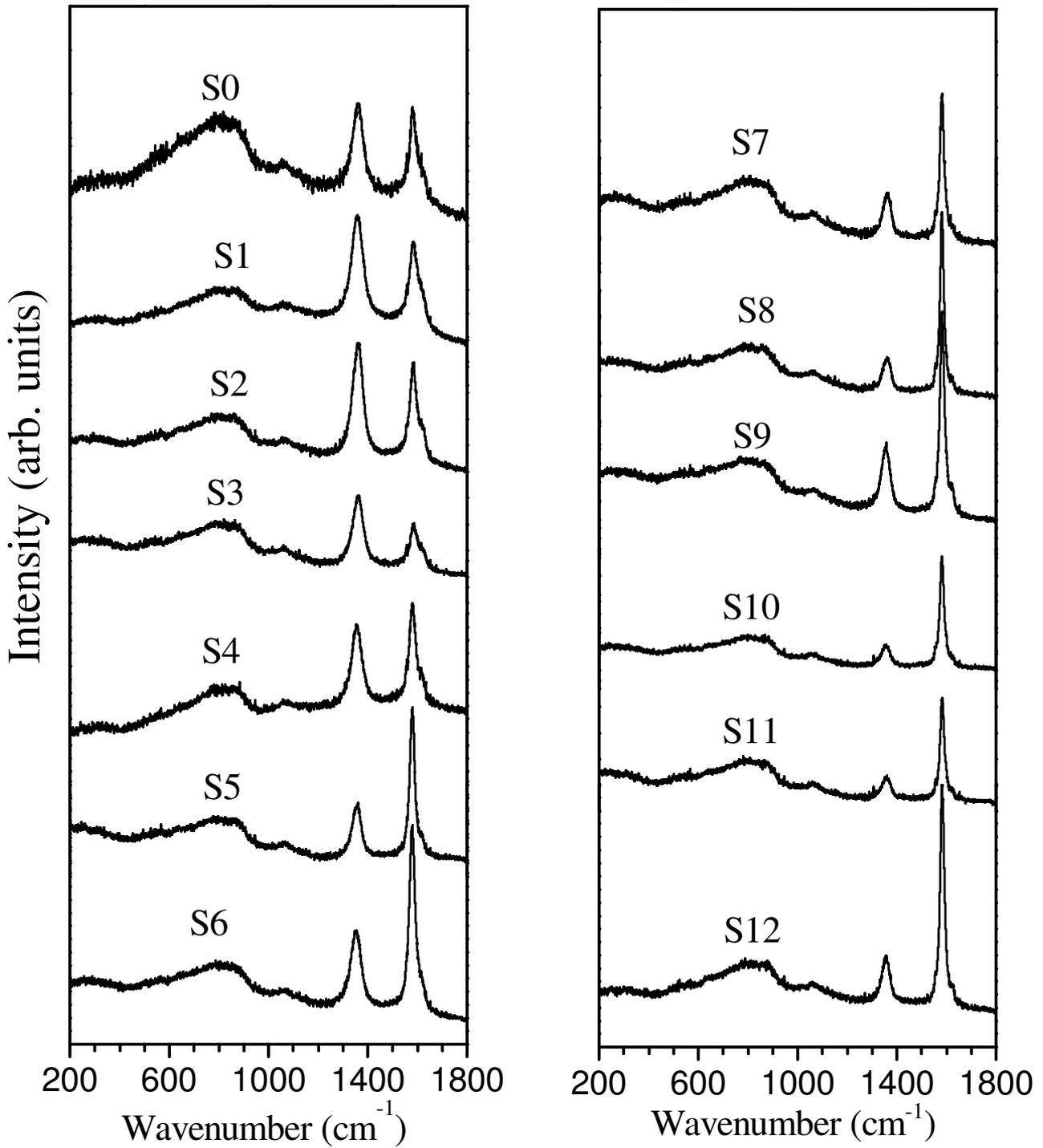

**Figure S4:** The peak position and peak width of the D and G peaks in the Raman spectra of the graphite samples, and the intensity ratio ($I_D/I_G$) of the D and G peaks. The irradiated samples are numbered as S0 to S11 in the order of decreasing neutron fluence seen by them; i.e., S0 and S11 have seen the maximum and minimum neutron fluence, respectively. Another un-irradiated sample for reference is assigned as sample number S12. For clarity, refer Fig. 1.

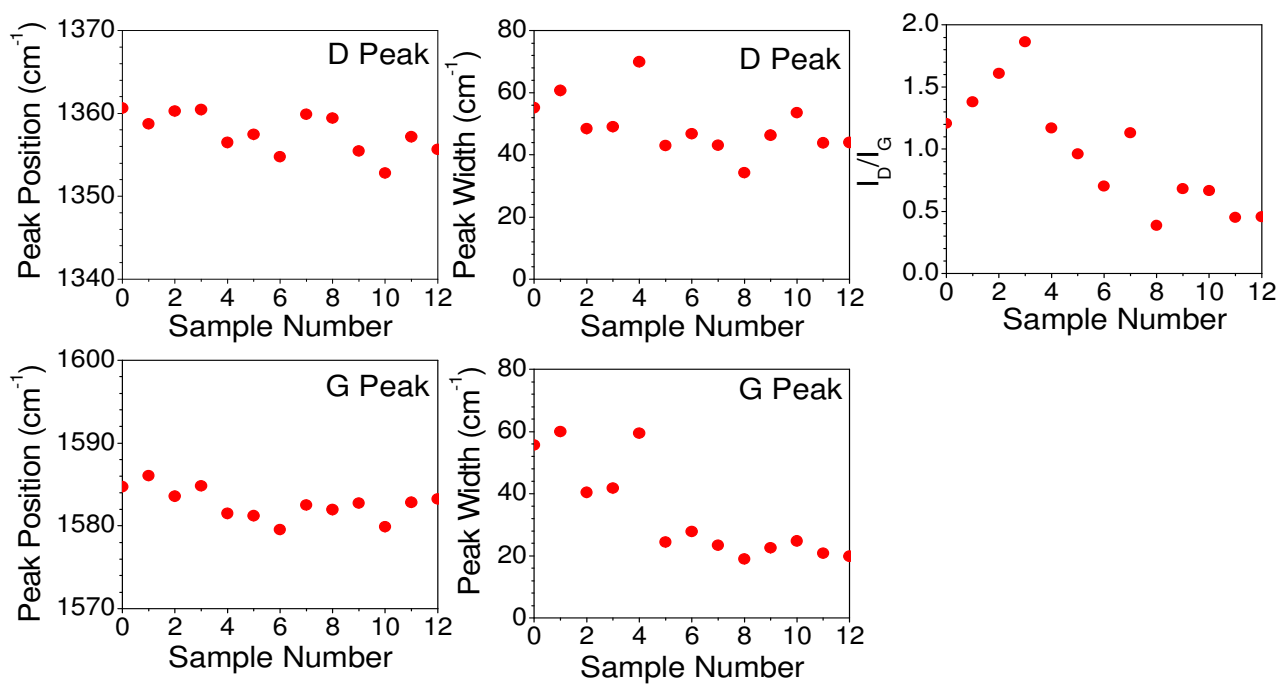

## IV. Differential Scanning Calorimeter Measurements

Quantitative information on extent of defects in the crystal lattice of graphite formed during irradiation of samples was determined by heating the irradiated samples in a Mettler Toledo differential scanning calorimeter (DSC-I). The DSC was calibrated for temperature and heat flow by melting high purity 'In' and 'Zn' standards. About 10-15 mg of the irradiated graphite samples in the form of powder were sealed in a 40 μl aluminium DSC pan. The samples were then heated at a rate of 5 °C/min from 25 to 400 °C under the flow of argon. The enthalpy changes for the process and the annihilation temperature were determined using STARe Software provided by M/s Mettler.

**Figure S5** gives the variation of enthalpy change for the annihilation of defects formed by the neutron irradiation for several graphite samples with different values of the adsorbed neutron dose. For each sample a single exothermic peak was observed in the temperature range 135 and 190 °C, which is due to annealing of defects. The plot of heat flux versus temperature for irradiated graphite samples with different doses is shown in **Fig. S5**. The values of heat released, temperature range and exposure doses for various graphite samples are given in **Table S1**. It is clear that the energy released during heating of various samples is related to the absorbed dose. The increase in absorbed dose results in increase in the formation of the defects in the graphite lattice.

**Figure S5:** Heat flow versus temperature for graphite sample irradiated with different fluence of neutron. The irradiated samples are numbered as S0 to S11 in the order of decreasing neutron fluence seen by them; i.e., S0 and S11 have seen the maximum and minimum neutron fluence, respectively. Another un-irradiated sample for reference is assigned as sample number S12. For clarity, refer Fig. 1.

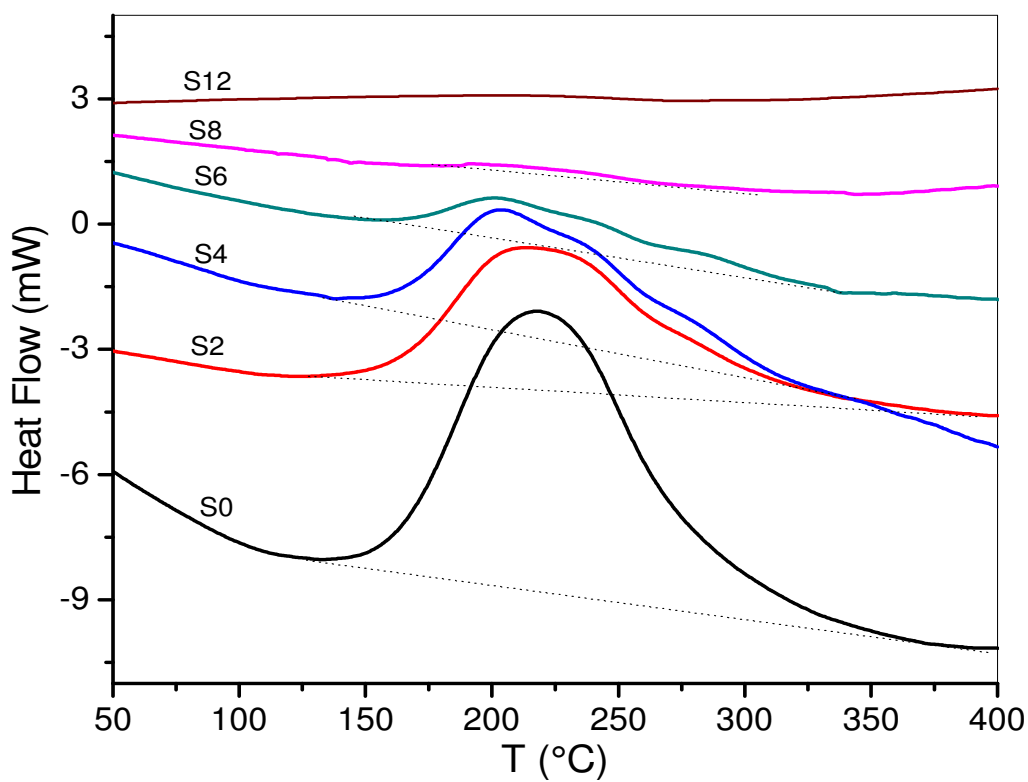

**TABLE S1.** The analysis of the data for various samples from differential scanning calorimeter experiment. The irradiated samples are numbered as S0 to S11 in the order of decreasing neutron fluence seen by them; i.e., S0 and S11 have seen the maximum and minimum neutron fluence, respectively. Another un-irradiated sample for reference is assigned as sample number S12. For clarity, refer Fig. 1.

| Sample No. | Onset Temperature for Wigner energy (°C) | Peak Temperature for Wigner energy (°C) | End Temperature for Wigner energy (°C) | Enthalpy Change on release of Wigner energy (J/g) |
|---|---|---|---|---|
| S0 | 135 | 218 | 380 | 117.90 |
| S2 | 135 | 219 | 390 | 82.30 |
| S4 | 139 | 207 | 375 | 75.51 |
| S6 | 150 | 205 | 350 | 17.00 |
| S8 | 170 | 212 | 294 | 1.52 |
| S12 | - | - | 0 | 0 |

## V. Small Angle X-ray Scattering Measurements

In order to characterize the pore size distribution in the samples we have performed small-angle X-ray scattering (SAXS) studies. The SAXS measurements were carried out using a Rigaku small angle goniometer mounted on rotating anode X-ray generator. Scattered X-ray intensity I(q) was recorded using a scintillation counter by varying the scattering angle 2θ where q is the scattering vector given by $4\pi.\sin(\theta)/\lambda$ and λ is the wavelength of incident ($CuK_\alpha$) X-rays. The intensities were corrected for sample absorption and smearing effects of collimating slits[3]. Data were recorded on graphite samples irradiated by neutrons of various neutron fluences.

**Fig S6(a)** shows the SAXS data displayed on log-log scale of three samples and the profile shape is similar for all other samples. The scattering profiles exhibit two distinct linear regions with change in the slope at a crossover point denoted by $q_o$, which indicates that there are two distinct types of pores in the material. This is in agreement with the earlier studies on irradiated graphite[4,5] which also showed the presence of large and small size pores. The slope of the lines in low-q region ($q < q_o$) varies in the range 3.50-3.60 indicating surface fractal behavior of larger pores with rough surfaces[6,7]. Since the linear region holds even at the smallest angle at which the scattering data are recorded, the larger pore dimension would be at least 170 nm.

For $q > q_o$, the slope is nearly 4.0 indicating smooth boundary for basic pores. From the crossover points, the estimated average dimension of the smaller pores varies in the range 10-14 nm. Thus, with change in the irradiation level, the overall morphology of microstructure was not affected, but a shift in $q_o$ or size of basic pores is observed. The effect of irradiation on the microstructure is strikingly different from the previous studies on nuclear graphite. Earlier study[4] showed mass fractal nature for larger size pores with fractally rough surfaces for small pores before irradiation. Following the irradiation, the surface fractal nature was completely suppressed. Whereas, in our study, the larger pores have not formed as mass fractals, but have surface fractal nature and the irradiation has not caused significant impact on the overall morphology of both small and large pores in the graphite. **Fig S6(b)** shows the variation of pore size with the neutron fluence, which indicates only a marginal variation.

**Fig. S6** (a) Small angle X-ray scattering profiles; lines are guide to the eye. (b) Variation of pore size with thermal neutron fluence. The irradiated samples are numbered as S0 to S11 in the order of decreasing neutron fluence seen by them; i.e., S0 and S11 have seen the maximum and minimum neutron fluence, respectively. Another un-irradiated sample for reference is assigned as sample number S12. For clarity, refer Fig. 1.

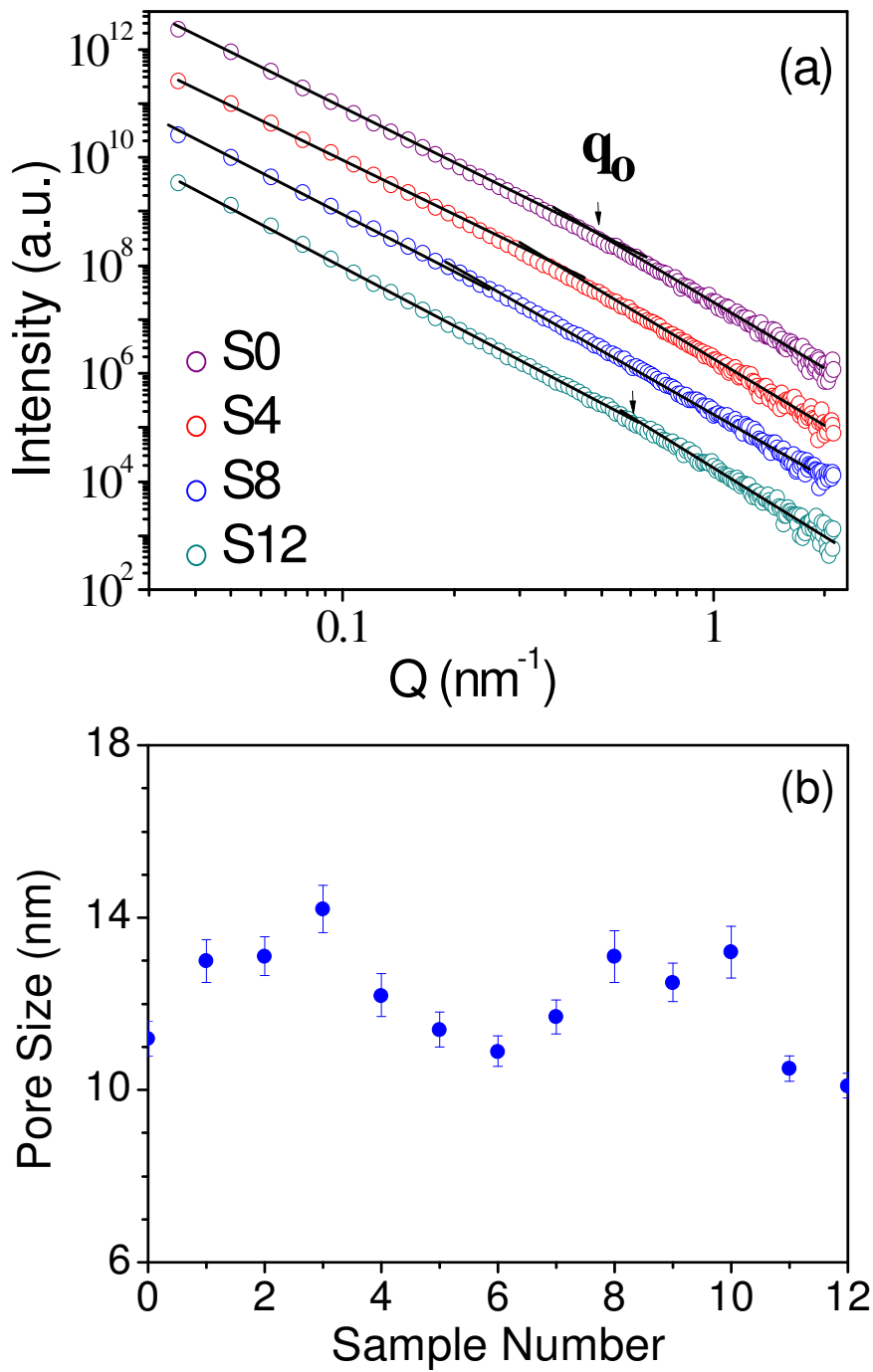

# VI. Structure and Dynamics of Graphite From Ab-initio Calculations

**TABLE S2**. Comparison between the calculated structural parameters of the perfect graphite structure[8] (P6$_3$/mmc) and the structures with a single Frenkel defect in 4×4×1 and 4×4×2 supercells of graphite. In the simulations, the defect structures were relaxed without assuming the overall hexagonal symmetry. The experimental values[8] of the a and c lattice parameters for graphite at room temperature are 2.462 Å and 6.707 Å respectively. We find that for the defect structure the *a*-lattice parameter changes only slightly, while the c-parameter increases substantially in qualitative agreement with experimental data in Fig. S2.

|  | a (Å) | b (Å) | c (Å) | α (°) | β (°) | γ (°) |
|---|---|---|---|---|---|---|
| Perfect structure unit cell (4 atoms) | 2.4650 | 2.4650 | 6.6930 | 90.0 | 90.0 | 120.0 |
| Perfect structure 4×4×1 supercell | 9.8600 | 9.8600 | 6.6930 | 90.0 | 90.0 | 120.0 |
| Perfect structure 4×4×2 supercell | 9.8600 | 9.8600 | 13.3860 | 90.0 | 90.0 | 120.0 |
| Single Frankel defect in 4×4×1 supercell | 9.8738 | 9.8738 | 7.1120 | 85.3 | 94.7 | 120.6 |
| Single Frankel defect in 4×4×2 supercell | 9.8665 | 9.8666 | 13.7893 | 87.5 | 92.5 | 120.3 |
| Two Frankel defects in 4×4×1 supercell | 9.8666 | 9.5814 | 7.7748 | 80.6 | 92.1 | 119.2 |

**Fig. S7** The calculated partial and total phonon density of states of graphite and its defect structures with a single Frankel defect in 4×4×1 and 4×4×2 supercells.

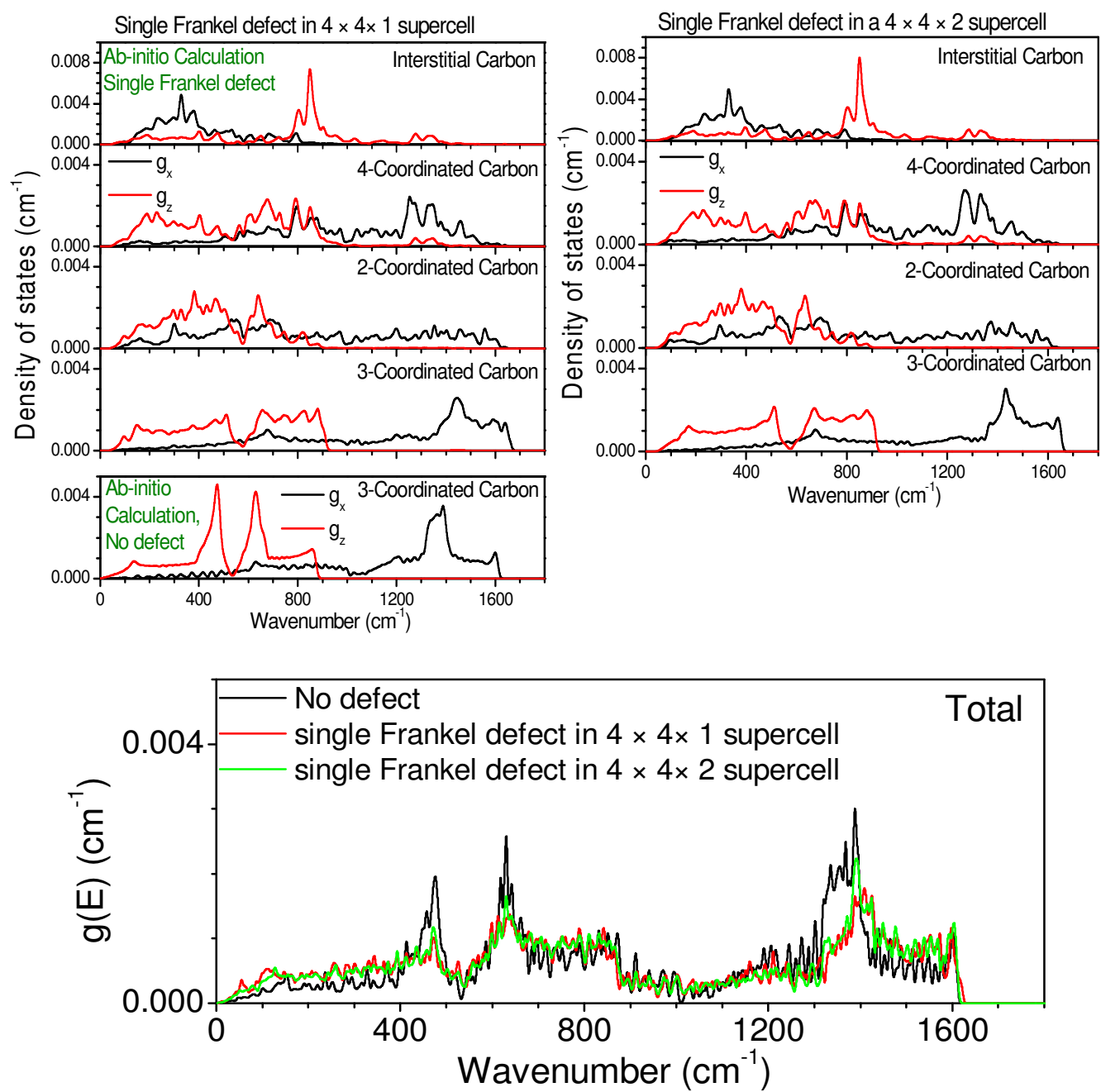

**Fig. S8** The total energy of 4×4×1 supercell (64 atoms) as a function of time with a single and two Frenkel defects from ab-initio molecular dynamics simulations.

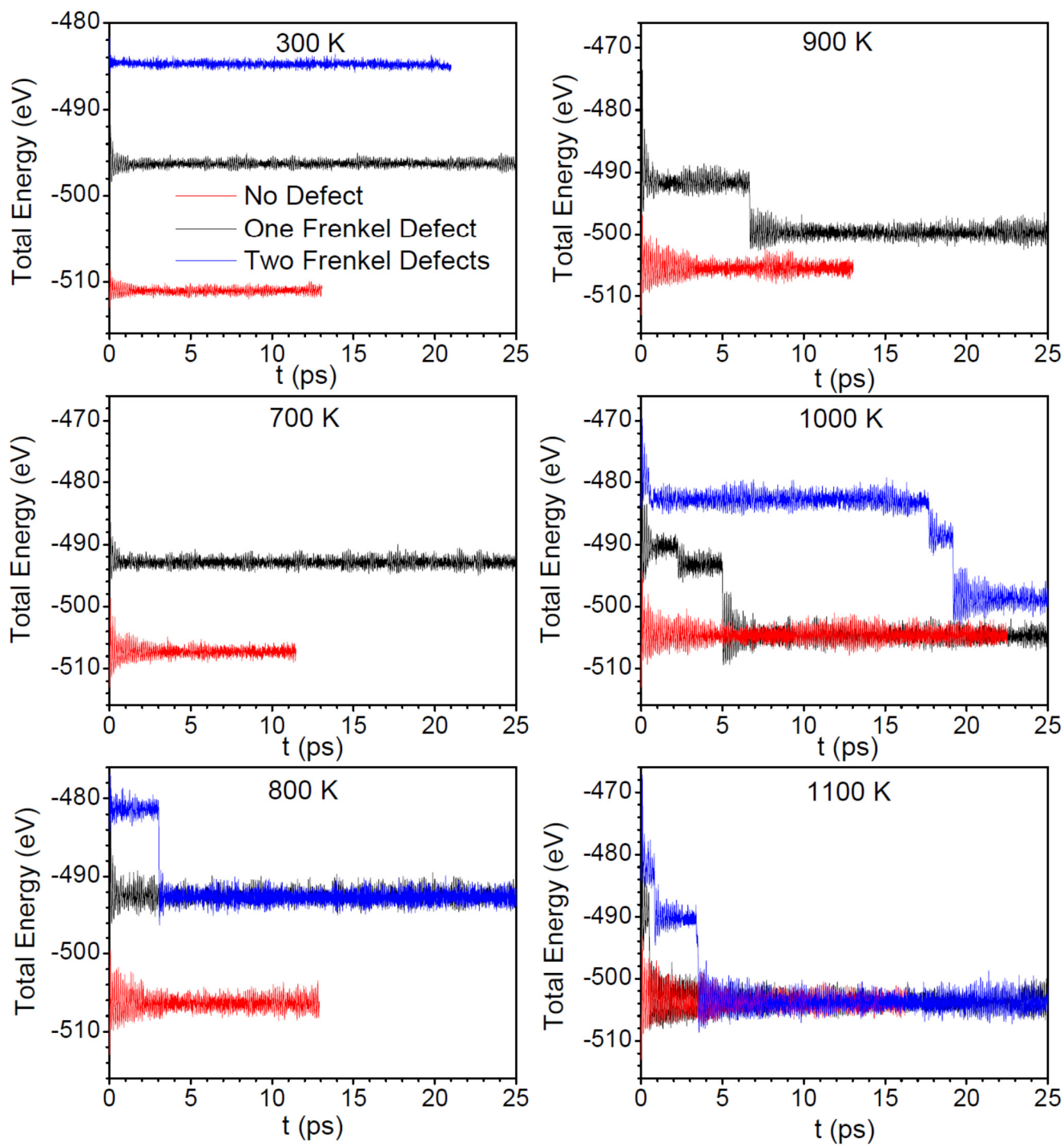